\documentclass[aps,prl,10pt,floatfix,twocolumn,longbibliography,superscriptaddress]{revtex4-2}

\usepackage[utf8]{inputenc} 
\usepackage[english]{babel}
\usepackage[normalem]{ulem}

\usepackage{amssymb,amsfonts,amsmath,mathrsfs,,physics}

\usepackage{graphicx}
\usepackage[usenames,dvipsnames]{xcolor}
\usepackage[colorlinks=true,citecolor=Blue,urlcolor=Blue,linkcolor=Blue]{hyperref}

\newcommand{\sP}[1]{\hat{S}^+_{#1}}
\newcommand{\sM}[1]{\hat{S}^-_{#1}}
\newcommand{\sz}[1]{\hat{S}^z_{#1}}

\begin{document}
\title{Finite steady-state current defies non-Hermitian many-body localization}

\author{Pietro Brighi}
\affiliation{Faculty of Physics, University of Vienna, Boltzmanngasse 5, 1090 Vienna, Austria}

\author{Marko Ljubotina}
\affiliation{Technical University of Munich, TUM School of Natural Sciences, Physics Department, 85748 Garching, Germany}
\affiliation{Munich Center for Quantum Science and Technology (MCQST), Schellingstr. 4, M\"unchen 80799, Germany}
\affiliation{Institute of Science and Technology Austria, Am Campus 1, 3400 Klosterneuburg, Austria}

\author{Federico Roccati}
\affiliation{Quantum Theory Group, Dipartimento di Fisica e Chimica Emilio Segr\`e, Universit\`a degli Studi di Palermo, via Archirafi 36, I-90123 Palermo, Italy}
\affiliation{Max Planck Institute for the Science of Light, 91058 Erlangen, Germany}

\author{Federico Balducci}
\affiliation{Max Planck Institute for the Physics of Complex Systems, N\"othnitzer Str.\ 38, 01187 Dresden, Germany}

\date{\today}


\begin{abstract}
    Non-Hermitian many-body localization (NH MBL) has emerged as a possible scenario for stable localization in open systems, as suggested by spectral indicators identifying a putative transition for finite system sizes. 
    In this work, we shift the focus to dynamical probes, specifically the steady-state spin current, to investigate transport properties in a disordered, non-Hermitian XXZ spin chain. Through exact diagonalization for small systems and tensor-network methods for larger chains, we demonstrate that the steady-state current remains finite and decays exponentially with disorder strength, showing no evidence of a transition up to disorder values far beyond the previously claimed critical point. Our results reveal a stark discrepancy between spectral indicators, which suggest localization, and transport behavior, which indicates delocalization. This highlights the importance of dynamical observables in characterizing NH MBL and suggests that traditional spectral measures may not fully capture the physics of non-Hermitian systems. 
    Additionally, we observe a non-commutativity of limits in system size and time, further complicating the interpretation of finite-size studies. These findings challenge the existence of NH MBL in the studied model and underscore the need for alternative approaches to understand localization in non-Hermitian settings.
\end{abstract}

\maketitle


Non-Hermitian (NH) Hamiltonians are a ubiquitous tool used to capture the dissipative nature of realistic quantum systems~\cite{AshidaAdvinPhys2020}. While a NH Hamiltonian does not fully characterize the dynamics of an open quantum system, it does when postselecting on the no-jump trajectories (no-click limit)~\cite{MingantiPRA2019}.
Recently, the interplay of many-body and NH physics has garnered significant interest due to the intriguing phenomena emerging in this setting.
These include interaction-induced and many-body skin effect~\cite{PoddubnyPRB2023,GarbeSciPost2024,GliozziPRL2024}, skin solitons~\cite{WangArXiv2024}, nonlinear exceptional points~\cite{FelskiArXiv2024,KwongArXiv2025}, and NH many-body localization (MBL)~\cite{Hamazaki2019NonHermitian}.

In Hermitian MBL, the presence of disorder can suppress transport, resulting in a non-thermal insulating phase~\cite{Gornyi2005Interacting,Basko2006Metal,Serbyn2013Local,Ros2015Integrals,Imbrie2016Diagonalization,*Imbrie2016Many,Abanin2019Colloquium,Sierant2025Many}.
The absence of transport and the breakdown of ergodicity have been used interchangeably as hallmarks of many-body localization, leading to the wide use of spectral indicators to detect MBL~\cite{Oganesyan2007Localization,Znidaric2008Many,Pal2010Many,Deluca2013Ergodicity,Luitz2015Many}.
In the context of open quantum systems, a large body of literature has shown that Hermitian MBL is destabilized by any bath with a continuous spectrum~\cite{Nandkishore2014Spectral,Gopalakrishnan2014Mean,Huse2015Localized,Johri2015Many,Fischer2016Dynamics,Levi2016Robustness,Medvedyeva2016Influence,Nandkishore2017Many,Luitz2017How,Everest2017Role,Wu2019Describing,Wybo2020Entanglement,Brighi2022Propagation}. However, as pointed out originally in Ref.~\cite{Hamazaki2019NonHermitian}, localization seems to persist in the no-click limit. 
This paradox can be resolved in some models, where it is possible to generalize to the many-body setting the similarity transformation that maps the non-Hermitian Hamiltonian to its Hermitian counterpart~\cite{Hatano1996Localization,*Hatano1997Vortex,*Hatano1998NonHermitian}. While the transformation is legitimate only with open boundary conditions, in the localized phase it is expected that the boundary conditions are irrelevant, thus enabling a mapping of the Hermitian MBL phase onto the NH MBL phase~\cite{Gliozzi2025NonHermitian}.

On the footsteps of the standard Hermitian case, many works have focused on the behavior of the spectra and eigenstates of NH Hamiltonians~\cite{Zhai2020Many,Suthar2022NonHermitian,Ghosh2022Spectral,Yamamoto2023Localization,Mak2024Statics,DeTomasi2024Stable,Akemann2025Two}---and sometimes on other indicators as the singular value decomposition~\cite{Roccati2024Diagnosing}. 
However, the \textit{exponential suppression of transport} with system size, defining feature of (Hermitian) localization~\cite{Anderson1958Absence,Gornyi2005Interacting,Basko2006Metal,DeRoeck2024Absence}, has not been investigated in the non-Hermitian setting.

In this work, we consider the disordered non-Hermitian XXZ chain and study the behavior of the spin current in the steady state by means of exact diagonalization (ED) for small system sizes and tensor-network methods (time-evolving block decimation, TEBD~\cite{Vidal2003}) for systems of up to $N=560$ spins. We find that the steady-state current is exponentially suppressed with the disorder strength in the range of parameters we investigate. Remarkably, the steady-state current does not display a non-analiticity as a function of the disorder strength, and remains finite up to values of disorder well beyond the putative critical point identified through spectral indicators at the center of the complex spectrum~\cite{Hamazaki2019NonHermitian}. This points to the absence of a true phase transition for observables (i.e.\ expectation values of Hermitian operators), suggesting that if localization takes place, it does so at higher disorder strengths. Our work highlights that traditional spectral measures of Hermitian MBL do not necessarily translate to its NH counterpart, where the role of states in the middle of the spectrum is less central to long-time dynamics.

\paragraph{Model.}
We study a one dimensional $N$-site spin chain governed by a non-Hermitian (NH) Hamiltonian
\begin{equation}
    \label{eq:H_eff}
    \hat{H}_\mathrm{eff} = \hat{H} - \imath \hat{\Gamma}
    = \hat{H} - \frac{\imath}{2}\sum_{j=1}^N\hat{L}^\dagger_j\hat{L}_j ,
\end{equation}
which describes the no-click limit of a Lindblad master equation $\dot \rho = -\imath[\hat H ,\rho ]+\sum_j\mathcal{D}[\hat L_j]\rho$, where $\mathcal{D}[\hat L]\rho = \hat L\rho\hat L^\dagger - (\hat L^\dagger \hat L\rho + \rho\hat L^\dagger \hat L)/2$ and $\rho$ is the density matrix of the $N$ spins. 
The Hermitian part corresponds to the disordered XXZ chain:
\begin{equation}
    \label{Eq:H_XXZ}
    \hat{H} = J\sum_{j=1}^N \left[ \frac{1}{2}\left( \sM{j} \sP{j+1} +  \sP{j} \sM{j+1}\right) + V \sz{j} \sz{j+1} + h_j \sz{j} \right],
\end{equation}
where $V$ is the interaction strength, $h_j$ are i.i.d.\ random variables uniformly distributed over the interval $[-h,h]$, $\hat{S}_j^{x,y,z}$ are the spin-1/2 operators acting on site $j$, and $\hat{S}^\pm = \hat{S}^x \pm \imath \hat{S}^y$. In the following we fix the energy scale by setting $J=1$. We will study the system under both open (OBC) and periodic boundary conditions (PBC).

The dissipative part is given by jump operators of the form $\hat{L}_j = \sqrt{\gamma}( \sM{j} + e^{\imath\theta} \sM{j+1})$ which give rise to non-reciprocal coupling in the no-click limit~\cite{Hanai2023Universality,Clerk2015Nonreciprocal,Wanjura2020Topological,Porras2019Topological}. 
The corresponding effective Hamiltonian is
\begin{multline}
    \label{Eq:H_eff_XXZ}
    \hat{H}_\mathrm{eff} = \sum_{j=1}^N \left[ \frac{1-\imath\gamma e^{\imath\theta}}{2} \sP{j} \sM{j+1} + \frac{1-\imath\gamma e^{-\imath\theta}}{2}\ \sM{j} \sP{j+1} \right.\\
    \left. \vphantom{\frac{1}{2}}+ V \sz{j} \sz{j+1} + \left( h_j -\imath \gamma \right) \sz{j} \right] -\imath\gamma \frac{N}{2}.
\end{multline}
The dissipative term induces non-reciprocal hopping amplitudes and, choosing $\theta = \pm\pi/2$, the left and right hopping amplitudes become real and imbalanced.
The effective Hamiltonian Eq.~\eqref{Eq:H_eff_XXZ} further acquires finite imaginary shifts proportional to the number of sites ($-\imath\gamma N/2$) and to the global magnetization ($-\imath\gamma \sum_j\sz{j}$), which ensure the physicality of its solutions, i.e.\ its eigenvalues lie in the lower part of the complex plane and thus the wavefunction norm is not exponentially amplified.
Finally, mapping spins-1/2 to fermions via a Jordan-Wigner transformation, one realizes that Eq.~\eqref{Eq:H_eff_XXZ} is the interacting version of the Hatano-Nelson model~\cite{Hatano1996Localization,*Hatano1997Vortex,*Hatano1998NonHermitian}.

In the following, aligning with the existing literature on the model, we will focus on the zero magnetization sector---which effectively removes the $-\imath \gamma \sum_j \sz{j}$ term from the Hamiltonian---and further drop the constant term $-\imath \gamma N/2$:
\begin{multline}
    \label{Eq:H_eff_XXZ_purged}
     \hat{H}_\mathrm{eff} = \sum_{j=1}^N \left[ \frac{1-\gamma}{2} \sP{j} \sM{j+1} + \frac{1+\gamma}{2}\ \sM{j} \sP{j+1} \right.\\
    \left. \vphantom{\frac{1}{2}}+ V \sz{j} \sz{j+1} + h_j \sz{j} \right] .
\end{multline}
This way, the complex spectrum becomes symmetric w.r.t.\ the real axis. 
We fix $\gamma=0.1$ and $V=1.1$ for all numerical simulations.

While the spectral properties of the Hamiltonian~\eqref{Eq:H_eff_XXZ_purged} have been investigated~\cite{Hamazaki2019NonHermitian,Zhai2020Many,Suthar2022NonHermitian,Ghosh2022Spectral,Yamamoto2023Localization,Mak2024Statics,DeTomasi2024Stable,Roccati2024Diagnosing,Akemann2025Two}, the presence or absence of transport in the NH case remains an open question. Clarifying this issue in the NH case is essential, especially because of the last developments in Hermitian MBL, showing that possibly the transition takes place at much stronger disorder values than originally thought~\cite{Suntajs2020Quantum,Panda2020Can,Abanin2021Distinguishing,Crowley2022Constructive,Morningstar2022Avalanches,Crowley2022Mean,Long2023Phenomenology}.
Crucially, in the NH setting there exists a \textit{steady state} coincinding with the eigenstate with the largest imaginary part of the corresponding eigenvalue~\footnote{Due to our conventions of $\imath$'s, the states with positive imaginary eigenenergies are amplified while those with negative imaginary eigenenergies are suppressed.}.
Therefore, \emph{a single eigenstate} determines transport at long times, in contrast with the Hermitian case where it takes contribution from all eigenstates.

\paragraph{Spin transport.}

As both the Hermitian and non-Hermitian Hamiltonians conserve the global magnetization $\hat{S}^z_\text{tot} = \sum_j\sz{j}$, we characterize the system's putative localization transition through the spin current.
In the Hermitian case the Heisenberg equation of motion for a local spin obeys a continuity equation and defines the current operator: $\partial_t \sz{j} = \hat{J}_j - \hat{J}_{j-1} = \big(\nabla\cdot\hat{J} \big)_j$, where $\nabla$ is the discrete derivative, and we introduce the current operator $\hat{J}_j := \frac{\imath}{2}(\sP{j}\sM{j+1} - \sM{j}\sP{j+1})$.
Including the NH terms, however, the Heisenberg equation is modified to $\imath \partial_t \sz{j} = \sz{j} \hat{H}_\mathrm{eff} - \hat{H}_\mathrm{eff}^\dagger \sz{j}$.
As a consequence, the continuity equation above changes, possibly making the definition of the current invalid in the NH case. 

Splitting the effective Hamiltonian~\eqref{Eq:H_eff_XXZ_purged} into Hermitian and anti-Hermitian components as in Eq.~\eqref{eq:H_eff}, one can observe that the anti-Hermitian part is proportional to the global current:
\begin{equation}
    \label{eq:Hamiltonian_current}
    \hat{H}_\text{eff} = \hat{H} - \imath\gamma\hat{J}_\text{tot}, \qquad
    \hat{J}_\text{tot} = \sum_{j=1}^N \hat{J}_j.
\end{equation}
It then follows that one can define a modified continuity equation for the NH case:
\begin{equation}
    \label{eq:continuity_eq_NH}
    \partial_t \sz{j} = \big(\nabla \cdot \hat{J}\big)_j + \gamma \big\{ \hat{J}_\mathrm{tot}, \sz{j} \big\}.
\end{equation}
This equation marks the difference between the NH and Hermitian cases: a non-local contribution to spin dynamics appears in what was a local continuity equation, as entailed by the anticommutator term. This additional term can be thought of as the source and sink induced by the local dissipation, and it is made non-local by the post-selection procedure~\cite{Kawabata2023}.

Before moving on, it is important to notice that the introduction of non-Hermiticity hinders the study of transport using states that are not eigenstates of the total magnetization $\sz{\mathrm{tot}}$~\cite{Mahoney2024Transport}. Indeed, if the initial state is an eigenstate of the magnetization, $\sz{\mathrm{tot}} \ket{\psi_0} = S_0 \ket{\psi_0}$, then at time $t$ $\ev{\sz{\mathrm{tot}}}{\psi(t)} = S_0 \braket{\psi(t)}{\psi(t)}$, as $[\hat{H}_\text{eff},\sz{\text{tot}}] = 0$. It is only the change in the norm of $\ket{\psi(t)}$ that causes the expectation value of $\sz{\mathrm{tot}}$ to change. Upon post-selecting, the norm of $\ket{\psi(t)}$ is restored to be 1, and the magnetization is truly conserved:
\begin{equation}
    \frac{\ev{\sz{\mathrm{tot}}}{\psi(t)}}{\braket{\psi(t)}{\psi(t)}} = S_0. 
\end{equation}
However, when the initial state $\ket{\psi_0}$ is \emph{not} an eigenstate of the magnetization, the NH dynamics does \emph{not} conserve the weight of the wavefunction in each separate magnetization sector. In fact, each sector is characterized by a different minimal decay rate, and in the infinite-time limit only one sector survives.

\begin{figure}
    \centering
    \includegraphics[width=\columnwidth]{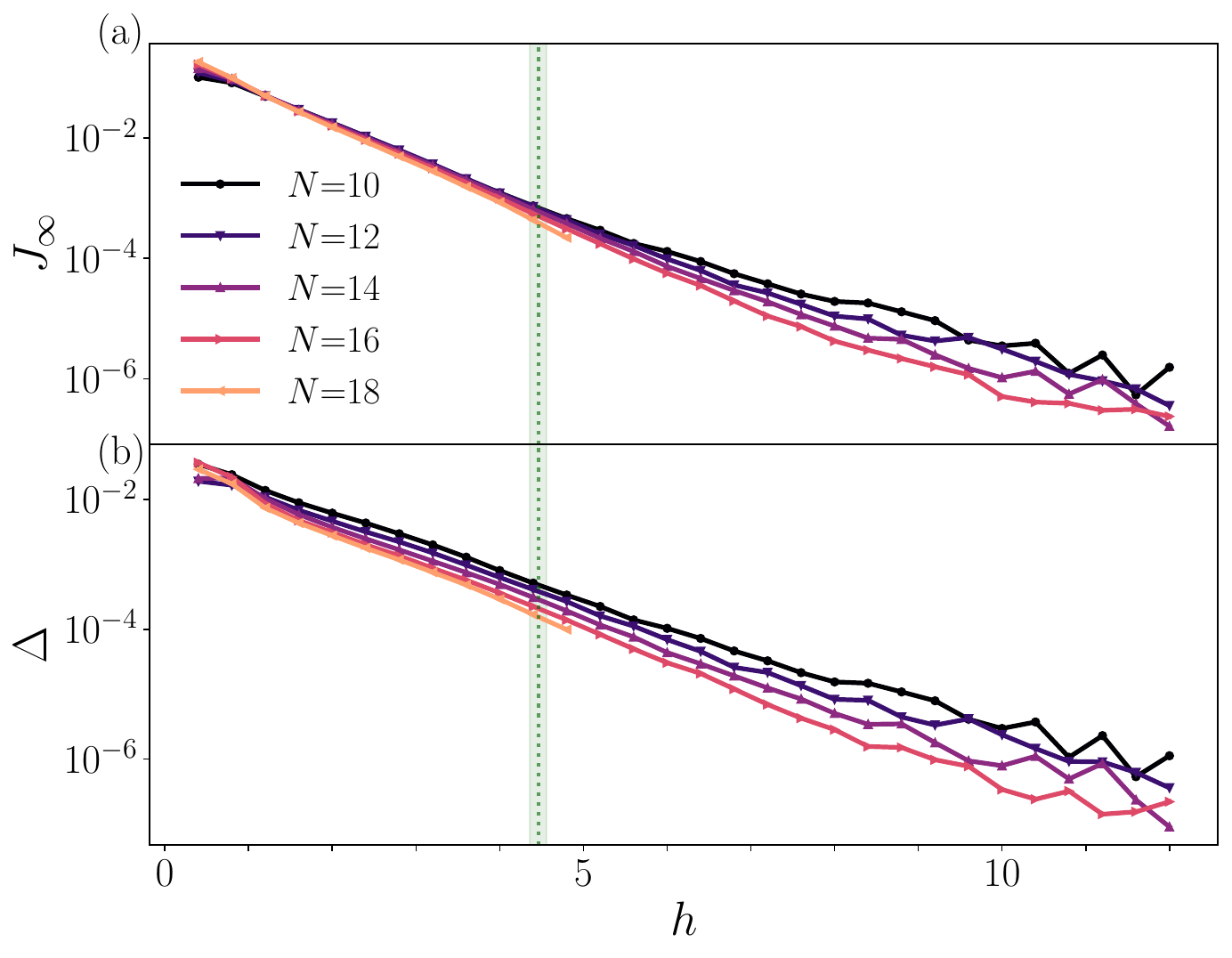}
    \caption{(a) Steady-state current $J_\infty$ obtained via exact diagonalization ($N \leq 16$) and via time evolution for long times ($N=18$). Around $h \simeq 4.5$ (green vertical stripe), a non-Hermitian many-body localization transition was previously claimed (see also Fig.~\ref{fig:fractions}). However, a weak current persists up to $h = 12$, signaling that the system is actually delocalized. (b) Gap to the first excited state $\Delta$, setting the (inverse) timescale at which the steady state is reached. The data is averaged over 15000 disorder realizations for $N\leq 16$, and over 3000 realizations for $N=18$.
    }
    \label{fig:current_gap}
\end{figure}

\paragraph{Numerical results.}
We first analyze the current in eigenstates using exact diagonalization (ED) for small system sizes up to $N=18$. Since the eigenvalues of $\hat{H}_\text{eff}$ are complex, the long-time behavior of any initial state is determined by the steady state. Therefore, instead of focusing on eigenstates in the middle of the spectrum, we will report results for the steady state. In particular, the relevant quantities of interest will be the imaginary part of its eigenvalue, which is equal to the steady-state current ($\Im E_1 = J_\infty$, see Eq.~\eqref{eq:Hamiltonian_current}), and the gap $\Delta = \Im E_1 - \Im E_2$ to the first excited state, whose inverse sets the timescale for the approach to the steady state. Notice that we are ordering the eigenvalues from highest to lowest imaginary part.

In general, the determination of extremal eigenvalues of a matrix is a simpler computational task w.r.t.\ the full ED of the matrix, or the extraction of eigenvalues in the middle of the spectrum~\cite{Pietracaprina2018Shift,Sierant2020Polynomially}. This would suggest that the steady-state properties of the model under consideration could be studied for larger system sizes than the ones usually accessible in the Hermitian counterpart. However, this expectation turns out not to be true for the model under consideration, and one needs to resort to the full ED of the Hamiltonian (see Supplement).

In Fig.~\ref{fig:current_gap}, we report the results for the total steady-state current $J_\infty$ (a), and the gap $\Delta$ to the second largest imaginary eigenvalue (b).  Our results indicate the absence of a transition at the critical value expected from indicators at the center of the spectrum $h_c \simeq 4.5$~\cite{Hamazaki2019NonHermitian,Mak2024Statics}, marked with a vertical green line in the figure. While the current is indeed suppressed as the disorder strength increases, it always remains finite in the range of parameters studied, going well beyond $h_c$. 

\begin{figure}
    \centering
    \includegraphics[width=\columnwidth]{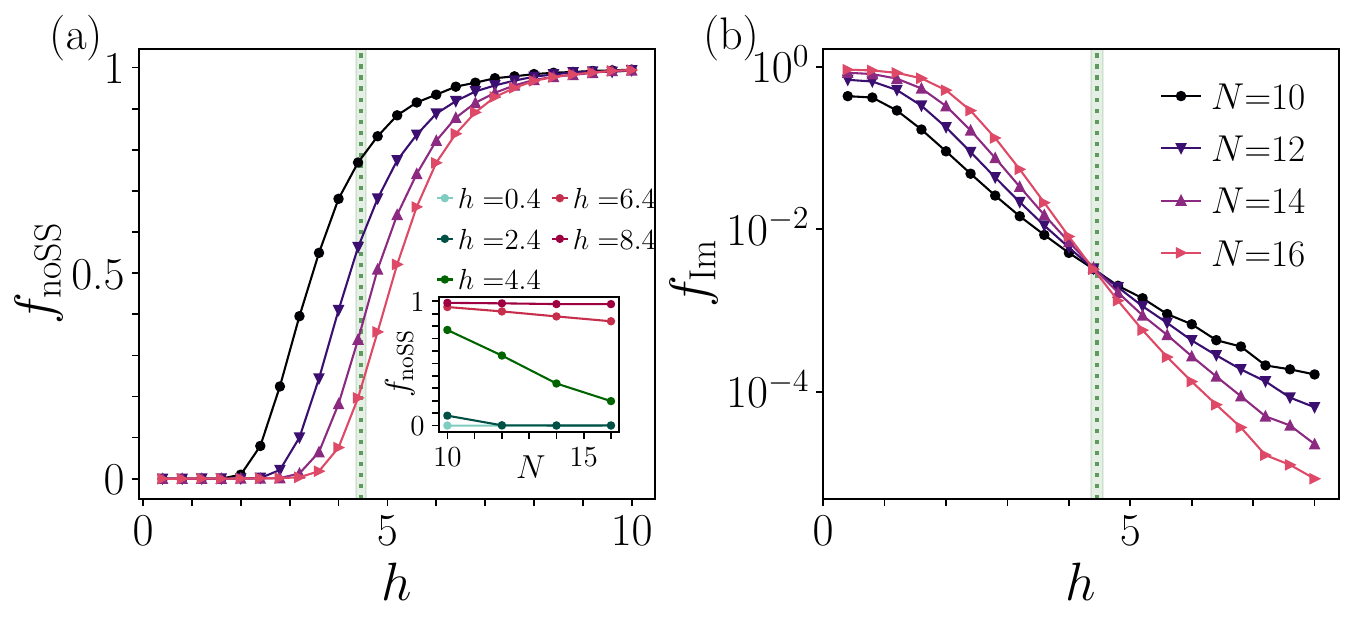}
    \caption{(a) The fraction $f_\mathrm{noSS}$ of disorder realizations with a completely real spectrum drifts towards larger values of the disorder strength $h$, as the system size is increased. Correspondingly, the spectrum becomes more and more delocalized. In the inset we show the scaling of $f_\mathrm{noSS}$ with system size at various values of $h$. (b) Considering instead the average fraction of eigenvalues with a nonzero imaginary part $f_\mathrm{Im}$, one might conclude that there is a NH MBL transition at the value $h_c \simeq 4.5$ (vertical green stripes). More details in the main text.}
    \label{fig:fractions}
\end{figure}

The global current decays as $J_\infty \propto e^{-\alpha h}$, with a rate $\alpha$ that depends only weekly on the system size---the precise behavior is hard to determine due to the small range of accessible system sizes. However, one can still confidently see that there is no sign of non-analiticity in $J_\infty$ as a function of $h$. This suggests that no transition occurs in the wide range of disorder strengths explored. These results are consistent with similar observations made in Hermitian models, which also showed exponentially suppressed transport but no signs of non-analiticity~\cite{Oganesyan2009Energy, Znidaric2018}. No non-analiticity is seen even in the dependence of the gap $\Delta \propto e^{-\beta h}$, with a different rate $\beta$. Also in this case, the dependence of $\beta$ on $N$ is too weak to draw any definite conclusion from the small system sizes one has access to.

To better understand why spectral indicators show the presence of a NH MBL transition while no sign is seen in the steady-state current $J_\infty$ or the gap $\Delta$, we plot the fraction of disorder realizations that yield a completely real spectrum, $f_\mathrm{noSS}$, in Fig.~\ref{fig:fractions}(a). One can see that the crossover is shifting towards larger values of $h$, thus larger system sizes allow for more delocalized steady states, as also shown by the system size scaling in the inset. The situation is somewhat reminiscent of the drifting of the finite-size critical point in Hermitian systems towards larger disorder values~\cite{Morningstar2022Avalanches,Crowley2022Constructive,Scoquart2024Role,Niedda2024Renormalization}. 

For comparison, in Fig.~\ref{fig:fractions}(b), we show a typical indicator used to detect the NH MBL transition: the average fraction of non-real eigenvalues in each realization of disorder, $f_\mathrm{Im}$ ~\cite{Hamazaki2019NonHermitian,Mak2024Statics}. Even if $f_\mathrm{Im}$ displays a crossover around $h_c \simeq 4.5$, and thus in the bulk of the spectrum many eigenvalues become purely real, the presence of nonzero imaginary eigenvalues well above $h_c$, testified by $f_\mathrm{noSS}$, means that at long times the dynamics will be delocalized and with a nonzero current flowing.

We complement the ED study with simulations of the system's dynamics using matrix product states (MPS) and a NH version of the time-evolving block decimation algorithm~\cite{Vidal2003}. As MPS algorithms are inefficient in periodic boundary conditions (PBC), we employ open boundary conditions (OBC). In OBC, however, the current is expected to vanish at long times, irrespective of the value of the disorder, as particles accumulate on one of the boundaries (the so-called \emph{NH skin effect}). To circumvent this issue, we simulate large chains $N=560$ and study the dynamics of the current at the center of the chain $\hat{J}_\text{mid} = \sum_{j=- \ell_0}^{\ell_0} \hat{J}_{N/2 + j}$, where the boundaries affect the system only after a sufficiently long timescale. Based on our ED results under PBC, we expect a finite current in the bulk before boundary effects kick in. As we show in the Supplement, before this timescale the current dynamics reaches a stable stationary plateau at $J_0(h)$ whose extent increases with system size. We then identify $j_0 = J_0/(2\ell_0)$ as a proxy for the current in the $N\to\infty$ steady state, and we compare it with the ED results, where the other order of limits is taken ($t\to\infty$ first).

\begin{figure}[t]
    \centering
    \includegraphics[width=0.99\linewidth]{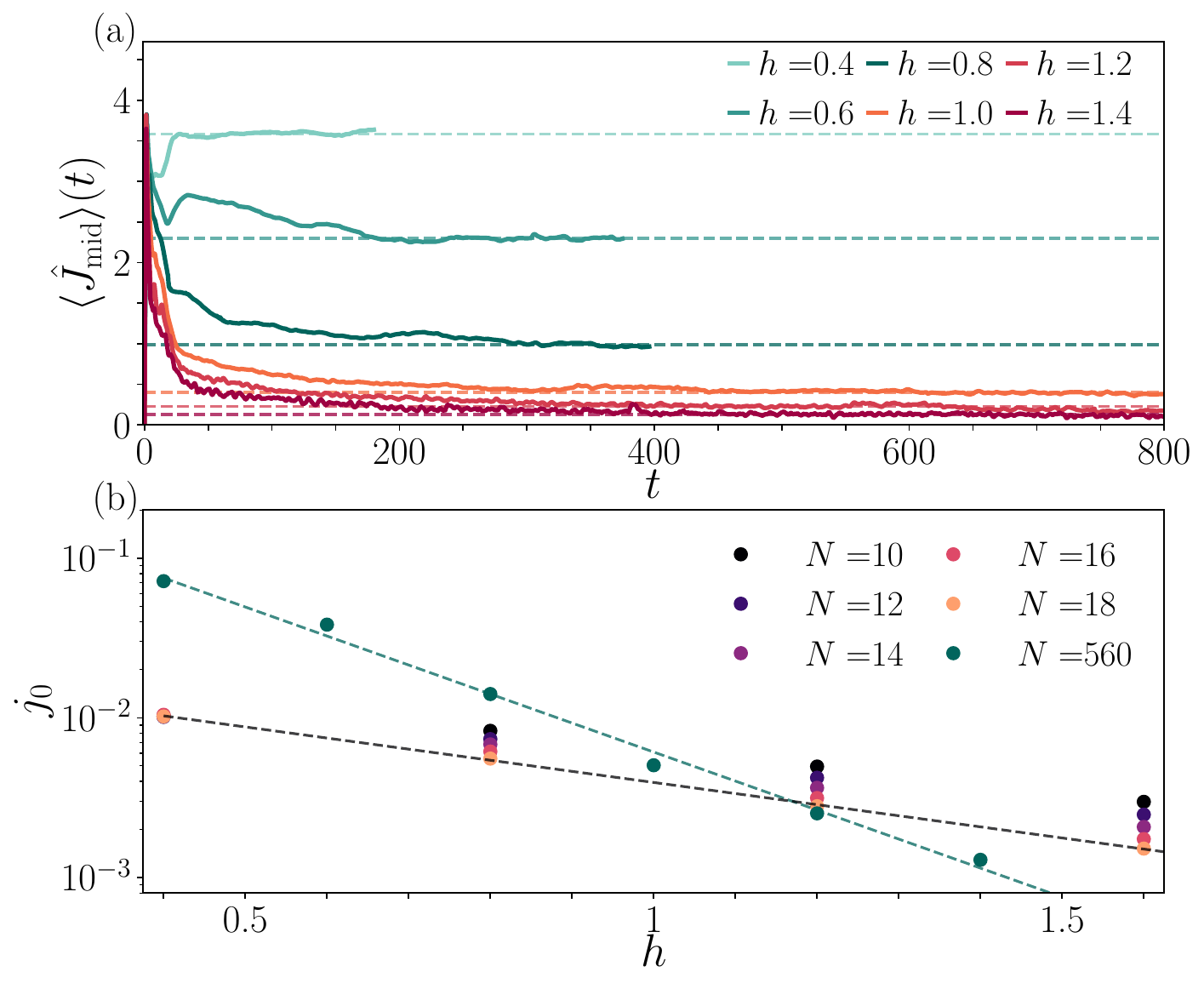}
    \caption{\label{Fig:Jmid Wcomp}
    (a): To avoid boundary effects, we study the current in the central $2\ell_0$ sites of a $N=560$ chain ($\ell_0 = 25,35,45$ for increasing disorder).
    The resulting $\langle \hat{J}_\text{mid}\rangle$ shows convergence in time to a plateau $J_0(h)$, before boundary effects eventually kick in.
    (b): The value $J_0$ is obtained by averaging the current within a large time window and we further evaluate the current per site $j_0 = J_0/\ell_0$ to be able to compare with exact diagonalization results.
    The value of $j_0$ obtained from dynamics also decays exponentially with $h$, but with a different slope than the one resulting from exact diagonalization.
    }
\end{figure}

In Fig.~\ref{Fig:Jmid Wcomp}, we report the bulk current dynamics obtained numerically from a N\'eel initial state $\ket{\psi_0} = \ket{\uparrow\downarrow\uparrow\downarrow\dots}$ using a large bond dimension $\chi = 768$. 
The current dynamics [Fig.~\ref{Fig:Jmid Wcomp}(a)] show a transient effect whose extent depends on the disorder strength $h$. At stronger disorder, dynamics are in general slower, leading to a longer transient phenomenon and reaching the plateau later. For the same reason, at weaker disorder boundary effects reach the central part of the chain at earlier times. Nevertheless, at the system size chosen these timescales allow to observe the formation of the current plateau, as clearly shown in the figure. As the current dynamics reaches a stationary condition, we use the average over a large time window to extract the value of the plateau $J_0$ (dashed lines). It is finally important to notice that the current reaches the plateau from above, thus meaning that spin flows (and the system is delocalized) not only asymptotically, but also at intermediate times.

The real-time simulation effectively gives access to values of the current in much larger systems than reachable by ED. One can then compare these results with the ones presented in Fig.~\ref{fig:current_gap}(a). In Fig.~\ref{Fig:Jmid Wcomp}(b), we show the behavior of the intensive bulk current $j_0 = J_0/(2\ell_0)$ and compare it to the ED results, where we take $j_0 = J_\infty/N$. Remarkably, the TEBD results confirm the exponential behavior of $j_0$ as a function of the disorder strength $h$, at least for small $h$. This indicates that the exponential scaling observed is robust, and possibly remains valid in the thermodynamics limit also at larger $h$. Interestingly, the slope of the exponential decay is different between ED and TEBD data. A possible explanation is that the slow drift in the exponent $\alpha$ seen at small system sizes leads to the larger $\alpha$ observed with TEBD. Another explanation could come from the two methods differing dramatically in the order of infinite-time and -system-size limits: while in ED one effectively takes $t\to\infty$ first by analyzing the behavior of eigenstates, in TEBD the opposite is true, as one first takes $N\to\infty$. The possibility of the two limits not commuting highlights yet again how the results on small systems obtained through ED must be interpreted with care.

\paragraph{Conclusion.}   
Our study challenges the existence of non-Hermitian many-body localization (NH MBL) in disordered spin chains by shifting the focus from spectral properties to dynamical transport. Using exact diagonalization and large-scale tensor-network simulations, we demonstrate that the steady-state spin current remains finite at all disorder strengths, decaying exponentially without any non-analyticity---contrary to expectations from spectral diagnostics. This suggests that the previously identified ``crossover'' based on mid-spectrum eigenstates does not manifest in physical observables, raising doubts about the stability of NH MBL in this setting.

Crucially, we find that the limits of infinite time and system size possibly do not commute, with spectral and transport probes yielding quantitatively different conclusions for finite systems. This underscores the necessity of dynamical approaches in studying non-Hermitian systems, where traditional Hermitian paradigms may fail. Our work calls for a reevaluation of NH MBL criteria, emphasizing that localization---if it exists---should be probed through observable quantities like transport, and not just spectral features. These results open new questions about the nature of dissipation-induced phenomena and the robustness of localization in open quantum systems.

\paragraph{Acknowledgments.}

F.~B.\ thanks Giuseppe de Tomasi and Oskar A. Pro\'sniak for discussion.
P.~B.\ acknowledges support by the Austrian Science Fund (FWF) [Grant Agreement No.~10.55776/ESP9057324].
This research was funded in whole or in part by the Austrian Science Fund (FWF) [10.55776/COE1].
The numerical simulations were performed using the ITensor library~\cite{ITensor} on the Vienna Scientific Cluster (VSC), and on the MPIPKS HPC cluster.
M.~L. acknowledges support by the Deutsche Forschungsgemeinschaft (DFG, German Research Foundation) under Germany's Excellence Strategy -- EXC-2111 -- 390814868. 
F.R.~ acknowledges support by the European Union-Next Generation EU with the project ``Quantum Optics in Many-Body photonic Environments'' (QOMBE) code SOE2024\_0000084-CUP B77G24000480006.


%

\onecolumngrid
\appendix
\begin{center}
    \vspace{1cm}

    \textbf{Supplemental Material for\\``Finite steady-state current defies non-Hermitian many-body localization''}
    \vspace{0.5cm}
    
    Pietro Brighi$^1$, Marko Ljubotina$^{2,3,4}$, Federico Roccati$^{5,6}$ and Federico Balducci$^7$
    \vspace{0.2cm}
        
    \small{\textit{$\text{ }^1$Faculty of Physics, University of Vienna, Boltzmanngasse 5, 1090 Vienna, Austria}}\\
    \small{\textit{$\text{ }^2$Institute of Science and Technology Austria, am Campus 1, 3400 Klosterneuburg, Austria}}\\
	\small{\textit{$\text{ }^3$Technical University of Munich, TUM School of Natural Sciences, Physics Department, 85748 Garching, Germany}}\\
	\small{\textit{$\text{ }^4$Munich Center for Quantum Science and Technology (MCQST), Schellingstr. 4, München 80799, Germany}}
    \small{\textit{$\text{ }^5$Quantum Theory Group, Dipartimento di Fisica e Chimica Emilio Segr\`e, Universit\`a degli Studi di Palermo, via Archirafi 36, I-90123 Palermo, Italy}}
    \small{\textit{$\text{ }^6$Max Planck Institute for the Science of Light, 91058 Erlangen, Germany}}
    \small{\textit{$\text{ }^7$Max Planck Institute for the Physics of Complex Systems, N\"othnitzer Str.\ 38, 01187 Dresden, Germany}}
\end{center}


\section{Finding the steady-state current of the non-Hermitian Hamiltonian}

In this work, we were concerned with finding the steady-state spin current for a non-Hermitian (NH) Hamiltonian. Because of the chosen convention of $\imath$'s, the steady state is approached dynamically as the real time goes to infinity, and it corresponds to the eigenstate of the Hamiltonian with largest imaginary part. 

As mentioned in the main text, finding extremal eigenvalues is usually a simpler computational task than finding the whole spectrum, or portions of it away from the spectral edges. This is one of the reasons why numerical studies of \emph{Hermitian} many-body localization are notoriously difficult to be carried out~\cite{Pietracaprina2018Shift,Sierant2020Polynomially,Sierant2025Many}. In the present study, one might think that the steady state is easily determined by looking for the extremal eigenvalue of the Hamiltonian $\hat{H}_\mathrm{eff}$ with largest imaginary part. This can be done with the help of recursive algorithms that need only sparse matrix multiplication. However, it is a rule of thumb that commonly used routines as Arnoldi need to converge within a Krylov sector of at least $O(50)$ eigenpairs, in order to yield a reliable estimate of the steady state. For the Hamiltonian under study, at large disorder only a handful of eigenvalues---if any at all---have a nonzero imaginary part, so the algorithm cannot resolve eigenvalues according to their imaginary part, and does not converge well.  Pictorially, the large number of real eigenvalues act as a ``wall'', preventing iterative algorithms to distinguish between eigenvalues that are degenerate in the imaginary part. Also, the presence of realizations with an entirely real spectrum makes it very difficult to discern the reason for Arnoldi not converging: a too small gap, or no gap at all. In conclusion, full diagonalization must be performed instead.

For the reasons above, we performed full exact diagonalization for system sizes up to $N=16$. To go to larger system sizes ($N=18$), we used instead that the real-time evolution operator $e^{-\imath \hat{H}_\mathrm{eff}t}$ acts as a projector on the steady state at long times (provided the norm of the evolved state is kept normalized). We explain here the procedure employed.

Let the expansion of the initial wavefunction in the energy (right) eigenbasis of $\hat{H}_\mathrm{eff}$ be
\begin{equation}
    \ket{\psi(0)} = \sum_n c_n \ket{r_n},
\end{equation}
so that the time-evolved state is
\begin{equation}
    \ket{\psi(t)} = \sum_n c_n e^{-\imath E_n t} \ket{r_n}.
\end{equation}
Let us label the eigenvectors so that $\gamma_1 \geq \gamma_2 \geq \dots$, where $E_n \equiv \epsilon_n + \imath \gamma_n$. We want to find out
\begin{enumerate}
    \item what is the value of $\gamma_1$;
    \item what is the value of the gap $\Delta = \gamma_1 - \gamma_2$.
\end{enumerate}
These questions can be answered by looking at how the time $t \to +\infty$ is reached. First, consider the norm
\begin{equation}
    \mathcal{N}_t := \sqrt{\braket{\psi(t)}{\psi(t)}}
    = \left[ \sum_n |c_n|^2 e^{2\gamma_n t} + \sum_n \sum_{m \neq n} c_m^* c_n e^{\imath (\epsilon_m^*-\epsilon_n)t + (\gamma_n+\gamma_m)t} \braket{r_m}{r_n} \right]^{1/2},  
\end{equation}
where $\braket{r_m}{r_n}$ need not be zero because of the non-Hermiticity of $H$. The dominant exponentials in the $t \to +\infty$ limit are those containing, in order of importance, $2 \gamma_1$, $\gamma_1+\gamma_2$ and $2\gamma_2$. Therefore, at large times
\begin{equation}
    \mathcal{N}^2_t \simeq |c_1|^2 e^{2\gamma_1 t} \left[ 1 + e^{- (\gamma_1 - \gamma_2)t} \left( \frac{c_1^* c_2}{|c_1|^2} e^{\imath (\epsilon_1^*-\epsilon_2)t} \braket{r_1}{r_2} + \mathrm{h.c.} \right) + \frac{|c_2|^2}{|c_1|^2} e^{-2(\gamma_1-\gamma_2) t} + \cdots \right].
\end{equation}
Now, the crudest approximation gives 
\begin{equation}
    \label{app:eq:log_norm}
    \log \mathcal{N}_t \simeq \log |c_1| + \gamma_1 t,
\end{equation}
which tells that the slope with which $\log \mathcal{N}_t$ explodes can be fitted to give $\gamma_1$. Equation~\eqref{app:eq:log_norm} also tells what is the time at which the asymptotic scaling sets in: for times $t < t_\star = -(\log |c_1|)/\gamma_1$ the norm is not less than 1, as Eq.~\eqref{app:eq:log_norm} would suggest, but rather it is of order 1 as the other right eigenstates also contribute. Therefore, starting from a random state $\ket{\psi(0)}$, the log-norm stays constant until it starts exploding at
\begin{equation}
    t_\star \simeq \frac{\log \dim \mathcal{H}}{\gamma_1} \simeq \frac{L \log 2}{\gamma_1}.
\end{equation}
Finally, Eq.~\eqref{app:eq:log_norm} predicts that the corrections to the scaling are exponentially small, provided $t > (\gamma_1-\gamma_2)^{-1}$. 

In order to extract also $\gamma_1 - \gamma_2$ from the limit, it is convenient to evaluate another quantity (above, the gap was regulating exponentially small corrections, which are difficult to extract reliably). A possibility is to measure the violation of the eigenvalue equation 
\begin{equation}
    \mathcal{E}_t := \frac{1}{\mathcal{N}_t^{1/2}} \left \| \hat{H}_\mathrm{eff} \ket{\psi(t)} - \frac{1}{\mathcal{N}_t}\ev{\hat{H}_\mathrm{eff}}{\psi(t)} \ket{\psi(t)} \right\|:
\end{equation}
at infinite time, indeed, $\ket{\psi(t)}$ becomes parallel to $\ket{r_1}$ and $\mathcal{E}_{t \to + \infty} = 0$; the factors of $\mathcal{N}_t$ are used in order to balance the exploding norm. This can be seen by defining $\ket*{\tilde{\psi}(t)} := \mathcal{N}_t^{-1/2} \ket{\psi(t)}$, so that it holds
\begin{equation}
    \mathcal{E}_t := \left \| \hat{H}_\mathrm{eff} \ket*{\tilde{\psi}(t)} - \ev*{\hat{H}_\mathrm{eff}}{\tilde{\psi}(t)} \ket*{\tilde{\psi}(t)} \right\|.
\end{equation}
Proceeding as above, one can find that at large times
\begin{equation}
    \mathcal{E}_t \simeq C e^{-\Delta t} + O \left(e^{-2 \Delta t} \right),
\end{equation}
where $C$ is a complicated combination of spectral data. However, even without specifying $C$, one is able to determine the gap $\Delta$ from a linear fit of $\log \mathcal{E}_t$. The gain in using $\mathcal{E}_t$ instead of $\mathcal{N}_t$ for determining the gap stems from the fact that it is the \emph{leading} term in $\mathcal{E}_t$ that decreases exponentially, not the subleading corrections.

To summarize, one can obtain the imaginary part of the eigenvalue with the largest imaginary part by time-evolving the system for a time $t \gtrsim \gamma_1^{-1}$ and looking at how the norm explodes. The second largest eigenvalue can be obtained by evolving for a time $t \gtrsim \Delta^{-1}$ and looking at how the error in the eigenvalue equation shrinks. This method requires only the application of the (sparse) matrix $\hat{H}_\mathrm{eff}$ repeatedly to the state $\ket{\psi(0)}$, but it is limited by the time needed for convergence (i.e.\ $\Delta^{-1}$). For this reason, in practice we could access only $h \lesssim 5$, as displayed in the main text.

\begin{figure}[t]
    \centering
    \includegraphics[width=0.99\linewidth]{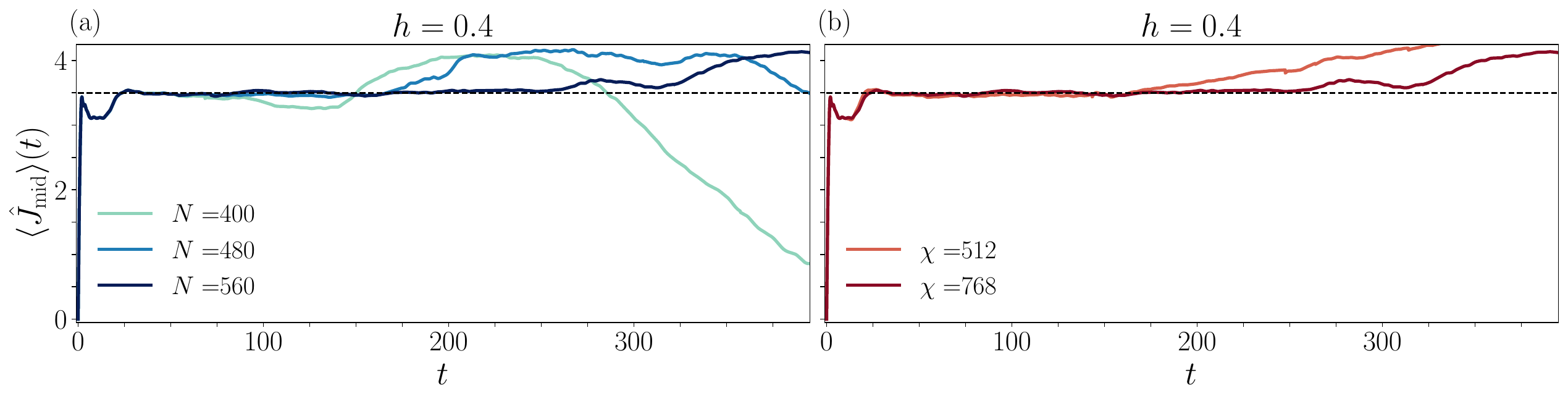}
    \caption{\label{Fig:J Comparison N}
    (a): We compare the mid current for various system sizes, showing how the current plateau expands as $N$ increases. 
    This suggests that the deviations from it are due to boundary effects and that the plateau corresponds to the asymptotic value as $N\to\infty$.
    (b): The value of the plateau is converged in bond dimension, ensuring the accuracy of our numerical simulations. We notice however that the extent of the plateau is affected by the choice of bond dimension.
    }
\end{figure}

\section{TEBD system size scaling}

In the main text we used TEBD to extract the asymptotic current in the bulk. To avoid boundary effects, we calculated the current in the central part of the chain, specifically in the region $[N/2-\ell_0,N/2 + \ell_0]$, with a varying $\ell_0$ depending on the value of $h$. As $h$ increases, the system becomes more localized, and on our timescales the boundaries of the system affect fewer sites, therefore at larger $h$ one can safely take larger $\ell_0$. To ensure that the current plateau we reported in the main text is indeed the asymptotic value in the thermodynamic limit, we performed system size scaling, as exemplified in Figure~\ref{Fig:J Comparison N}(a) for the case $h = 0.4$. The scaling clearly highlights that the current plateau at $J_0(h=0.4)\approx3.5$ extends in time as $N$ increases until eventually boundary effects kick in, suggesting that it corresponds to the value in the thermodynamics limit $N\to\infty$.

Besides confirming the validity of our analysis through system size scaling, we also compare results for different bond dimensions $\chi$.
In the main text we reported results for $\chi=768$, and here we compare them to $\chi=512$ to ensure their convergence.
As we show in Figure~\ref{Fig:J Comparison N}(b), increasing the bond dimension from $\chi=512$ to $\chi=768$ does not change the value of the current plateau $J_0$.
Therefore, the results we show in the main text are converged in bond dimension.
However, notice that the extent of the plateau is shorter at smaller bond dimensions: this further suggests that the observed plateau is the correct asymptotic one.

\end{document}